\documentclass[twocolumn,showpacs,preprintnumbers,amsmath,amssymb,prl,superscriptaddress]{revtex4}

\usepackage{epstopdf}

\usepackage{graphicx}
\usepackage{multirow}
\usepackage{comment}
\usepackage{color}

\begin{document}

\title{Tuning Magnetic Coupling in Sr$_2$IrO$_4$ Thin Films with Epitaxial Strain}

\author{A. Lupascu}
\author{J. P. Clancy}
\author{H. Gretarsson}
\author{Zixin Nie}
\affiliation{Department of Physics, University of Toronto, 60 St.~George St., Toronto, Ontario, M5S 1A7, Canada}
\author{J. Nichols}
\author{J. Terzic}
\author{G. Cao}
\author{S.~S.~A. Seo}
\affiliation{Department of Physics and Astronomy, University of Kentucky, Lexington, Kentucky 40506, USA}
\author{Z. Islam}
\author{M. H. Upton}
\author{Jungho Kim}
\author{D. Casa}
\author{T. Gog}
\author{A. H. Said}
\affiliation{Advanced Photon Source, Argonne National Laboratory, Argonne, Illinois 60439, USA}
\author{Vamshi M. Katukuri}
\affiliation{Institute for Theoretical Solid State Physics, IFW Dresden, Helmholtzstrasse 20, 01069 Dresden, Germany}
\author{H. Stoll}
\affiliation{Institute for Theoretical Chemistry, Universit\"{a}t Stuttgart, Pfaffenwaldring 55, D-70569 Stuttgart, Germany}
\author{L. Hozoi}
\author{J. van~den~Brink}
\affiliation{Institute for Theoretical Solid State Physics, IFW Dresden, Helmholtzstrasse 20, 01069 Dresden, Germany}

\author{Young-June Kim}
\email{yjkim@physics.utoronto.ca}
\affiliation{Department of Physics, University of Toronto, 60 St.~George St., Toronto, Ontario, M5S 1A7, Canada}

\date{\today}

\begin{abstract}
We report x-ray resonant magnetic scattering (XRMS) and resonant inelastic x-ray scattering (RIXS) studies of epitaxially-strained $\rm Sr_2IrO_4$ thin films. The films were grown on $\mathrm{SrTiO_3}$ and $\mathrm{(LaAlO_3)_{0.3}(Sr_2AlTaO_6)_{0.7}}$ substrates, under slight tensile and compressive strains, respectively. Although the films develop a magnetic structure reminiscent of bulk $\rm Sr_2IrO_4$, the magnetic correlations are extremely anisotropic, with in-plane correlation lengths significantly longer than the out-of-plane correlation lengths. In addition, the compressive (tensile) strain serves to suppress (enhance) the magnetic ordering temperature $T_N$, while raising (lowering) the energy of the zone boundary magnon. Quantum chemical calculations show that the tuning of magnetic energy scales can be understood in terms of strain-induced changes in bond lengths.

\end{abstract}

\maketitle

%%%%%%%%%%%%%%%%%%
%% Introduction %%
%%%%%%%%%%%%%%%%%%

%\input{intro2.tex}

The physics of strong spin orbit coupling (SOC) in condensed matter systems has been drawing increased interest in recent years. In particular, iridates have emerged as interesting model systems in which novel magnetism arises due to entangled spin and orbital degrees of freedom \cite{BJKim2009,JKim2012a}. A prototypical example is the layered $\rm Sr_2IrO_4$ (SIO), in which Ir$^{4+}$ ($5d^5$) ions form a square lattice. Due to the strong SOC, the Ir orbital moment is not quenched in this compound, and the local magnetic moment is described by a spin-orbit coupled $\mathrm{j_{eff}=1/2}$ instead of the usual spin only value commonly observed in lighter transition metals \cite{Crawford1994,Cao1998,BJKim2008,BJKim2009, morettisalaPRL2014}. A main reason for the strong interest in $\rm Sr_2IrO_4$ is that its magnetic properties are strikingly similar to those of the parent compounds of cuprate superconductors \cite{JKim2012a,Fujiyama2012}, raising the possibility that unconventional superconductivity could be realized in this system  by doping \cite{Wang2011,Watanabe2013}. Although superconductivity has not been realized so far, doping studies have shown that structural details, such as the Ir-O-Ir angle and the Ir-Ir distance, are important factors  to consider when studying the magnetic properties of $\mathrm{Sr_2IrO_4}$ \cite{Qi2012,Cava1994,Rao2000,Gatimu2012, PhysRevB.52.9143_shimura_1995, klein&terasaki_2008, cosio-castenada_2007}. However, since doping may also affect the charge concentration in addition to the crystal structure, an alternative means to tune the structure is necessary to elucidate the structure-property relation in $\mathrm{Sr_2IrO_4}$.

One of the most promising approaches to the structural tuning of oxide materials is strain engineering, accomplished by growing thin films on substrates with varying degrees of lattice mismatch. This method has been successfully used to study 3d and 4d transition metal oxides. For example, it was found that strained thin film cuprates show an increase in superconducting $T_c$ \cite{Locquet1998}. Strain can also be used to tune the properties of ferroelectrics such as  $\mathrm{SrTiO_3}$ \cite{Haeni2004} and $\mathrm{BaTiO_3}$ \cite{Choi2004}. The study of iridate thin films is still in its early stages, and most studies to date have focused on the structural and electronic properties of thin film $\rm Sr_2IrO_4$\cite{JSLee2012,Rayan2013,Nichols2013,Nichols2013_aaxis}. Rayan Serrao \emph{et al.} studied  $\rm Sr_2IrO_4$  films with various thicknesses grown on $\mathrm{SrTiO_3}$ substrates (slight tensile strain), and reported that thinner samples exhibit smaller c/a ratio \cite{Rayan2013}. They also suggested that electronic anisotropy is reduced in thinner samples based on their structural and x-ray spectroscopic data. Nichols \emph{et al.} grew $\rm Sr_2IrO_4$ films on substrates with varying degrees of strains, ranging from highly compressive to highly tensile \cite{Nichols2013}. They found that the optical absorption peak shifts to higher energies under tensile strain. Until recently, the magnetic properties of these thin film samples have been largely unexplored \cite{PhysRevB.89.035109_miao_2014}.

In this Letter, we report complementary x-ray resonant magnetic scattering (XRMS) and resonant inelastic x-ray scattering (RIXS) studies on $\rm Sr_2IrO_4$ thin film samples epitaxially grown on $\mathrm{SrTiO_3}$ (STO) and $\mathrm{(LaAlO_3)_{0.3}(Sr_2AlTaO_6)_{0.7}}$ (LSAT) substrates, which have been chosen to provide tensile and compressive strain, respectively. The most surprising result is that the magnetic ordering temperature $T_N$ is found to be suppressed (enhanced) in samples with compressive (tensile) strain. This observation is somewhat counter-intuitive, since both RIXS experiments and quantum chemical calculations predict magnetic interaction to strengthen when the in-plane lattice constant  shrinks. The magnetic coupling energy scale determined from the zone-boundary magnon energy increases (decreases) under compressive (tensile) strain, which is also well reproduced in calculated values for magnetic exchange constants. We argue that the observed behaviour of $T_N$ could be accounted for by the subtle change in inter-layer magnetic coupling due to in-plane strain. Our findings illustrate that the magnetic properties of $\rm Sr_2IrO_4$ are highly sensitive to the effects of epitaxial strain.

%%%%%%%%%%%%%%%%%%
%% Experimental %%
%%%%%%%%%%%%%%%%%%

The $\mathrm{Sr_2IrO_4}$ thin films, 20 unit-cells thick $\mathrm{(\approx 50\: nm)}$, were grown using pulsed laser deposition, as described in Ref.~\cite{Nichols2013}. The films were deposited on two different substrates: STO (100) (SIO-STO), with a 0.45 \% nominal tensile strain, and LSAT (100) (SIO-LSAT), with a 0.45 \% nominal compressive strain. The XRMS measurements were conducted at the Advanced Photon Source (APS) using beamline 6-ID-B. The data were collected at the Ir $\mathrm{L_3\:(11.215\:keV)}$ absorption edge. A graphite (008) polarization analyzer was used to select outgoing photon polarization for XRMS.
Ir L$_3$-edge RIXS measurements were carried out using the MERIX spectrometer on beamline 30-ID-B at the APS. Measurements were performed using a spherical (2 m radius) diced Si (844) analyzer and a channel-cut Si (844) secondary monochromator to give an energy resolution (FWHM) of 45 meV. The RIXS data was collected in horizontal scattering geometry, with a scattering angle close to $\mathrm{2 \theta = 90^{\circ}}$, to minimize the background contribution from elastic scattering.  To maximize the signal, measurements were performed near glancing incidence, with an angle of incidence $\mathrm{\alpha<1^{\circ}}$, for both the thin films and bulk sample.

%%%%%%%%%%%%%%%%%%%%%%%%%%
%% Results and Analysis %%
%%%%%%%%%%%%%%%%%%%%%%%%%%

%++++++++++++++++++++++++%
%%XRMS &structure       %%
%++++++++++++++++++++++++%

%XRMS results

%INSERTING FIG 1
\begin{figure}[!!h]
\includegraphics[width=3.5 in]{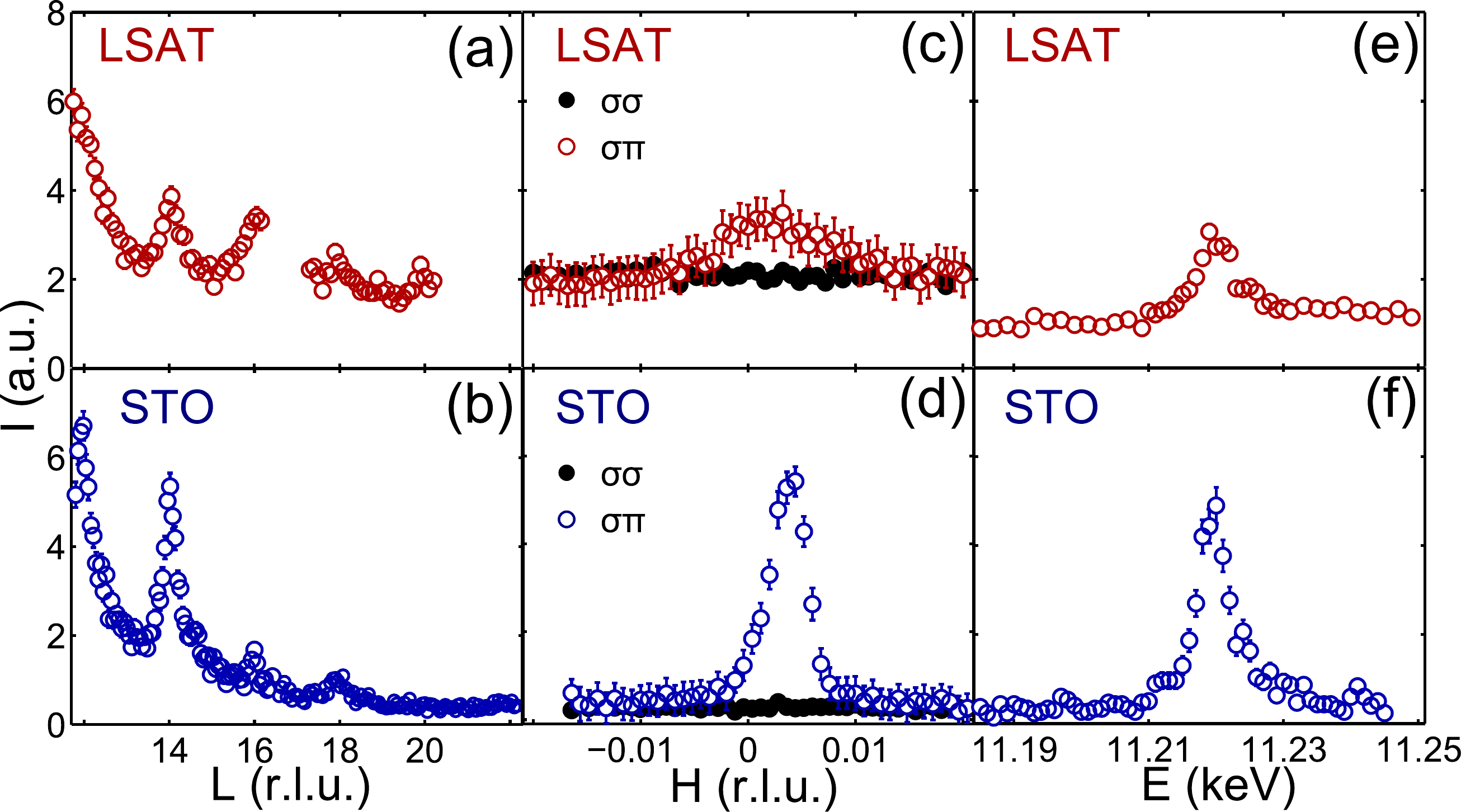}
\caption{\label{fig1} L-scan profile of magnetic x-ray diffraction along (0,1,L) for SIO-LSAT (a) and SIO-STO (b). Polarization dependence for SIO-LSAT (c) and SIO-STO (d), in the $\mathrm{\sigma\sigma}$ channel (filled circles) and the $\mathrm{\sigma\pi}$ channel (empty circles). Incident energy dependence for SIO-LSAT (e) and SIO-STO (f). All data in (c)-(f) were obtained at $T=5$~K at the $\mathrm{(0,1,14)}$ reflection.}
\end{figure}	

The tetragonal structure of bulk $\mathrm{Sr_2IrO_4}$ gives rise to two distinct magnetic domains. The first domain is characterized by magnetic reflections observed at $\mathrm{(1,0,4n+2)}$ and $\mathrm{(0,1,4n)}$, with integer n. The second domain gives rise to peaks at  $\mathrm{(0,1,4n+2)}$ and $\mathrm{(1,0,4n)}$ ~\cite{ Dhitak_2013_PhysRevB.87.144405}. With an x-ray beam spot much larger than the size of the magnetic domains, we expect to observe both types of domains and thus find peaks at all the even L positions of $\mathrm{(1,0,L)}$ and $\mathrm{(0,1,L)}$. Magnetic peaks have also been observed for odd $L$ values when small magnetic field is applied or when doping occurs on the Ir sites~\cite{BJKim2009,2012PhRvB.86v0403C,PhysRevB.89.054409_clancy_2014}.

%INSERTING FIG 2
\begin{figure}[!h]
\includegraphics[width=3 in]{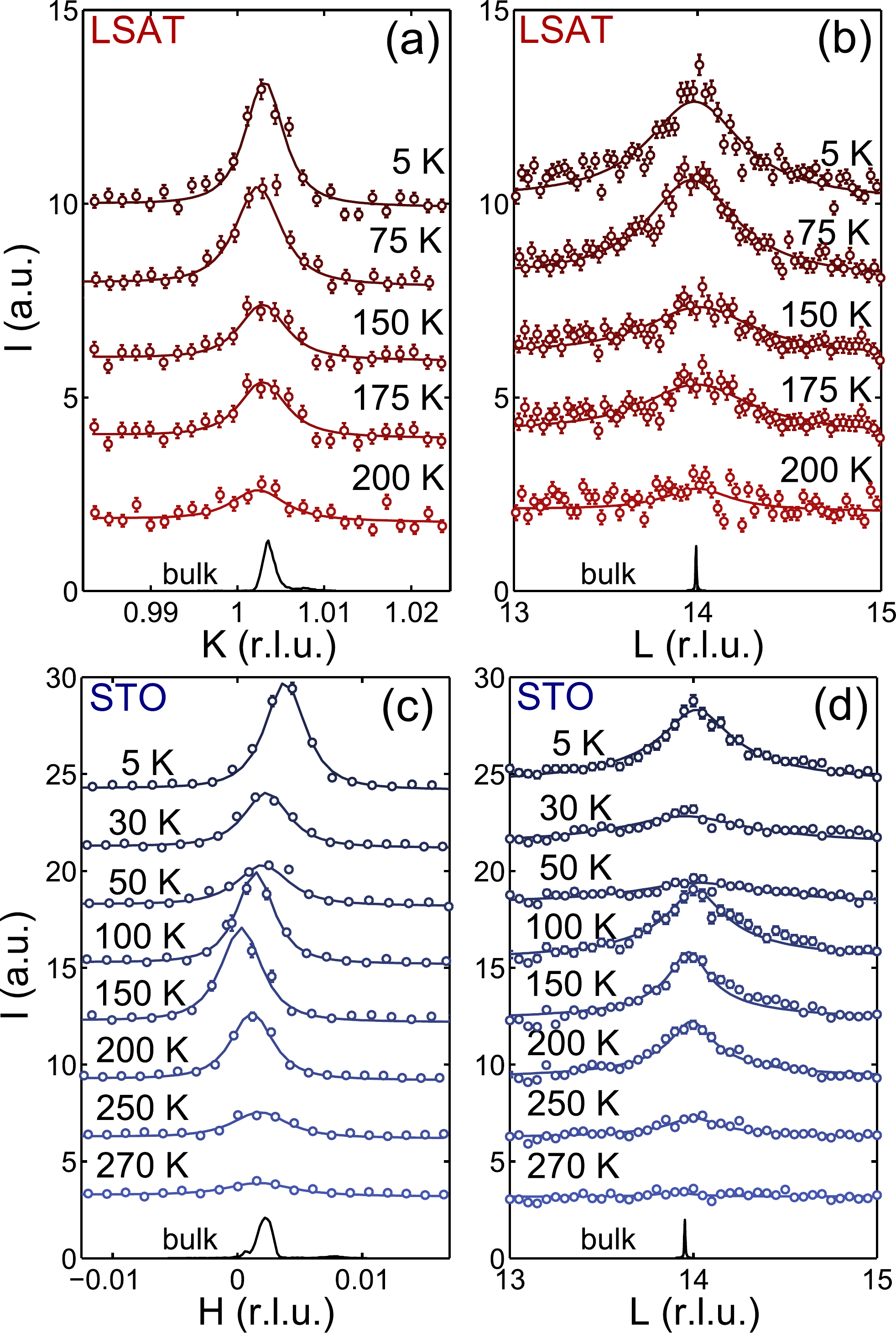}
\caption{\label{fig2} SIO-LSAT: Scans of the $\mathrm{(0,1,14)}$ magnetic peak along the K (a) and L-direction (b). SIO-STO: Scans of the $\mathrm{(0,1,14)}$ magnetic peak along the H (c) and L-direction (d). The baseline of the profiles in (a-d) is shifted. The solid lines through the data points are the results of the fit described in the text. The solid black lines depict the H, K, and L-scans of the (1,0,18) magnetic peak for bulk $\mathrm{Sr_2IrO_4}$ (the peak intensity is re-scaled and the L-scan is shifted from 18 to 14). }
\end{figure}

As shown in Fig.~\ref{fig1}(a)-(b), we have observed magnetic peaks at even $L$ positions in both thin films, which is consistent with the magnetic structure of bulk $\mathrm{Sr_2IrO_4}$ at zero-field. The magnetic nature of these peaks is illustrated in Fig. 1(c)-(d),which presents two scans obtained in different polarization channels. The energy dependence of the $\mathrm{(0,1,14)}$ magnetic peaks is shown in Fig. 1(e)-(f). The peaks resonate at an incident energy of 11.219 keV. These observations are all very similar to those in bulk crystals of $\mathrm{Sr_2IrO_4}$.

To further characterize the magnetic ordering of the films, we have studied the  $\mathrm{(0,1,14)}$ peak as a function of temperature. For SIO-LSAT, the K-scans (constant H and L) and L-scans (constant H and K) are shown in Fig.~\ref{fig2} (a) and (b). Similar sets of scans are plotted in Fig.~2 (c)-(d) for SIO-STO. Both in-plane scans (H and K scans) are quite similar and only one of these are shown. Unlike the SIO-LSAT film, which shows monotonous temperature dependence, the SIO-STO shows anomalous temperature dependence in the $\mathrm{30-100~K}$ range. These anomalies are presumably due to a variation in strain with temperature caused by structural transitions in the STO substrate. STO adopts several distinct structural phases at low temperatures: with a cubic to tetragonal transition at ($\mathrm{T=110\: K}$) and several other transitions at lower temperatures~\cite{Lytle1964}. The temperature dependence illustrates the close relation between strain and magnetism in this material. However, a further quantitative characterization of the SIO-STO film structure at low temperatures is beyond the scope of this paper, and we will only focus on $\mathrm{T>110\: K}$ data here.

\begin{figure}[!!h]
\includegraphics[width=3 in]{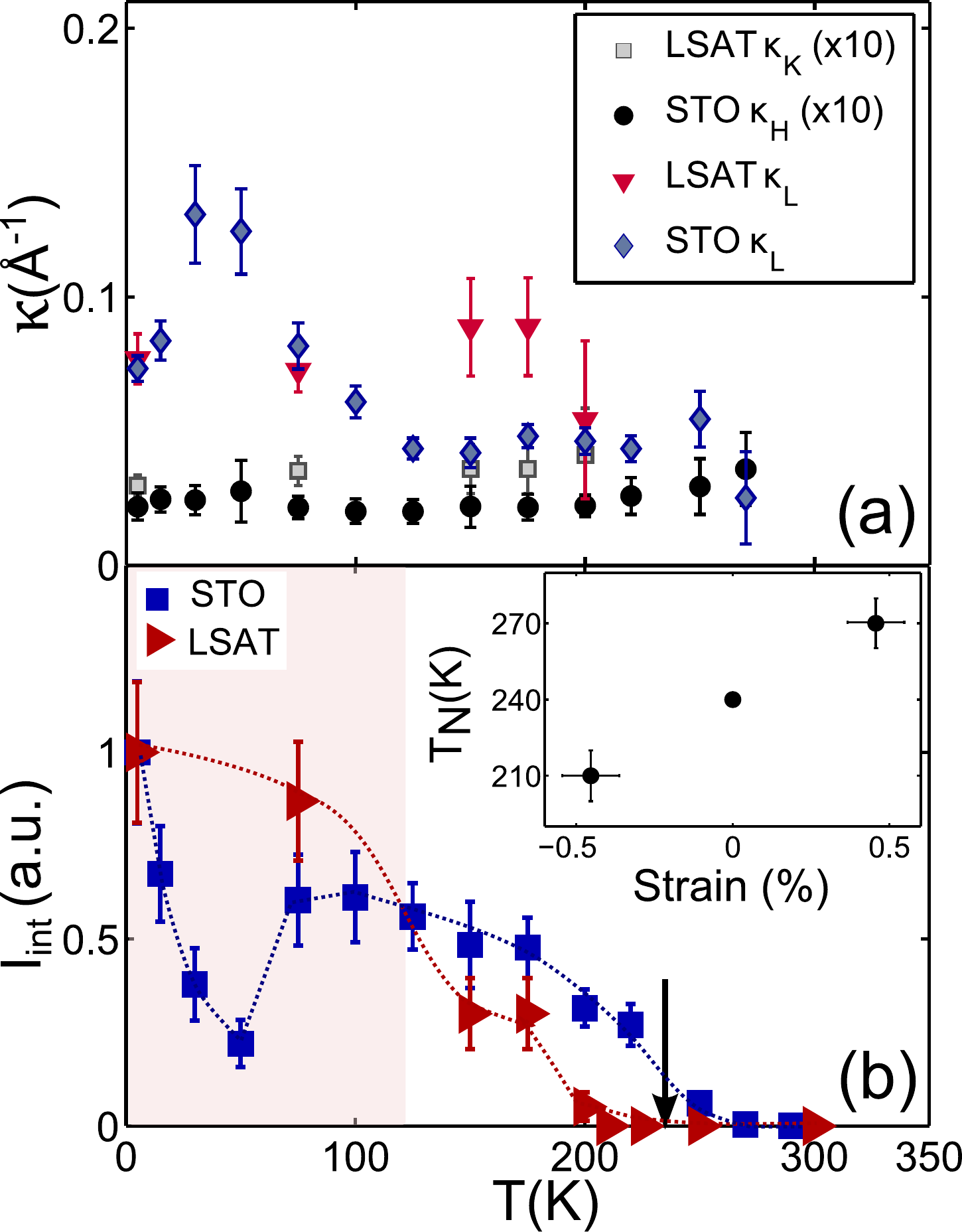}
\caption{\label{fig3} (a) The HWHM (fitting parameter $\kappa$) is plotted as a function of temperature, along the H and L-direction for SIO-STO, and along the K and L-direction for SIO-LSAT. (b) The change in integrated intensity with temperature for the $\mathrm{(0,1,14)}$ reflection, for SIO-LSAT (red triangles) and SIO-STO (blue squares). The black arrow indicates the  magnetic transition temperature of SIO bulk, and the red and blue dotted lines are provided as a guide to the eye. The background colouring depicts the structural transitions of the STO substrate. Anomalous temperature dependence below 110 K is due to structural transitions in the STO substrate, and will not be discussed here. The inset shows the magnetic transition temperature as a function of strain.}
\end{figure}

The peak profiles were fitted with a n-th power Lorentzian function: $I(q)=I_{max}[(q-q_0)^2/{(\zeta_n\kappa)^2}+1]^{-n}$, where $\zeta_n=\sqrt{2^{1/n}-1}$ is a constant set to keep $\kappa$ the half-width at half maximum (HWHM). The parameter $q$ is either H, K, or L, and $q_0$ is the respective peak position. The fitting procedure is detailed in the Supplemental Material~\cite{supplmat}. The fitting results are presented in Fig.~\ref{fig3}(a). The widths of the film magnetic peaks are approximately 2-3 times broader than the (1,0,18) bulk magnetic peak along the H and K-direction, and an order of magnitude broader along the L-direction, as shown in Fig. 2. The magnetic correlation length, $\xi$, can be estimated by inverting $\kappa$: $\mathrm{\xi=\kappa^{-1}}$. Both the tensile and the compressive strain films show similar magnetic correlation lengths, with a considerable anisotropy characterized by very small correlation lengths along the c-axis, $\mathrm{\xi \approx 10-20\:\AA}$ (approximately one unit-cell) and much larger correlation lengths in the ab-plane, $\mathrm{\xi \approx 300-400\:\AA}$. Such a short-range magnetic correlation along L is in stark contrast to the magnetic ordering in bulk  $\mathrm{Sr_2IrO_4}$, which develops into a full 3D long-range order ~\cite{Fujiyama2012}.

%INSERTING FIG 3

In Fig.~3(b) we present the temperature dependence of the integrated intensity for SIO-LSAT (red triangles) and SIO-STO (blue squares). The magnetic transition temperatures in the films are very different from that of bulk ($T_N\mathrm{=240\:K}$)~\cite{BJKim2009, Jackeli2009}. For the compressive strain (LSAT), the transition temperature is lower than the bulk, with a $T_N$ of only 210 K. The tensile strain (STO) has an increased $T_N$ of 270 K. This trend is illustrated in the inset of Fig. 3(b), which shows the transition temperature as a function of strain.

%++++++++++++++++++++++++%
%%RIXS & Theory         %%
%++++++++++++++++++++++++%

\begin{figure}[!!h]
\includegraphics[width=2.8 in]{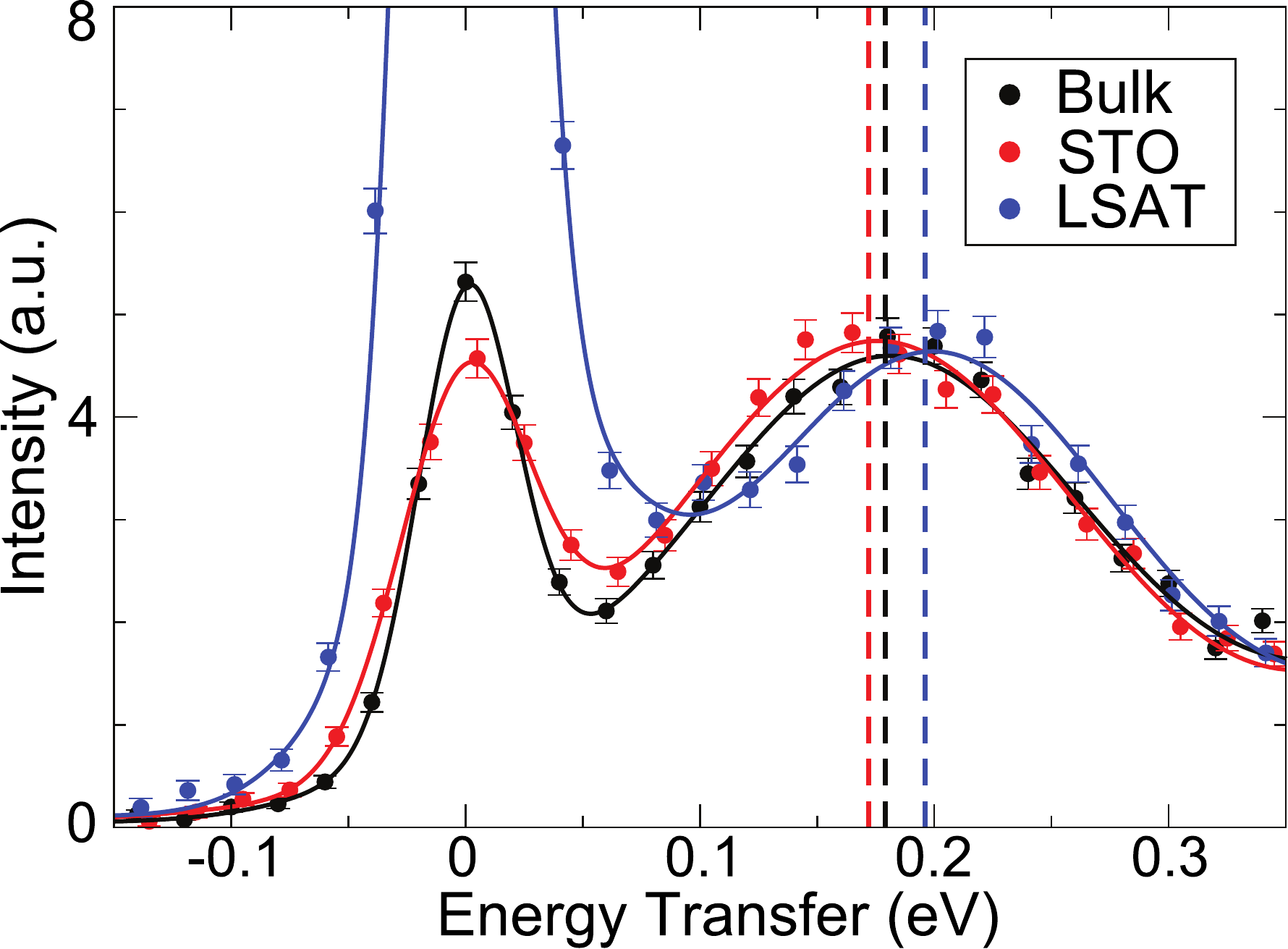}
\caption{\label{fig4} Epitaxial strain effect on the low-lying magnetic excitations of Sr$_2$IrO$_4$.  A comparison of Ir L$_3$-edge RIXS spectra collected at room temperature for bulk SIO, SIO-LSAT, and SIO-STO demonstrates that compressive (tensile) strain significantly raises (lowers) the energy of the ($\pi$,0) zone boundary magnon.  The solid lines represent Gaussian fits to the data (described in the Supplemental Material~\cite{supplmat}). The dashed vertical lines represent the fitted values for the magnon energies.}
\end{figure}

%RIXS
Representative RIXS measurements performed on bulk SIO, SIO-STO, and SIO-LSAT are presented in Fig.~\ref{fig4}. They highlight the strain effect on the magnon mode in $\mathrm{Sr_2IrO_4}$, which is observed at energy transfers of $\sim$170--200 meV \cite{JKim2012a}. The spectra presented were collected at the ($\pi$,0) zone boundary position, where the magnon energy is at its maximum. The RIXS excitation spectra for the SIO, SIO-STO and SIO-LSAT samples, including the spin-orbit exciton mode at $\mathrm{(\pi,0)}$ and $\mathrm{(pi/2,pi/2)}$ zone boundary wave-vectors, can be found  in Fig. 2 of the Supplemental Material~\cite{supplmat} \nocite{Huang1994,qc_NNs_degraaf_99,CuO2_dd_hozoi11,molpro_brief,ECP_Stoll_2,GBas_molpro_2p,113Ir_bogdanov_2012,Os227_bogdanov_12,ECP_Ir_Stoll,FuentealbaJPB,J_ligand_fink94,J_ligand_fink94,NOCI_J_oosten96,J_ligand_calzado03,Ir214_BHKim_2012,PhysRevB.89.035143}. The strain dependence of the zone boundary magnon energy, E$_{(\pi,0)}$, is very different from that of T$_N$, with compressive strain (SIO-LSAT) driving the magnon to higher energies, and tensile strain (SIO-STO) driving it lower.  Quantitative values for E$_{(\pi,0)}$, extracted from multi-Gaussian data fits, are provided in Table I.

%INSERTING FIG 4

On a qualitative level, the strain-induced tuning of magnetic energy scales in SIO can be understood as follows. The application of compressive epitaxial strain results in a reduction of the in-plane lattice parameters ({\it a} and {\it b}) and an enlargement of the out-of-plane lattice parameter ({\it c}). The magnetic exchange interactions between neighbouring Ir ions are very sensitive to bond geometry, so a decrease of Ir-O/Ir-Ir bond lengths will enhance the interaction strength and vice-versa. Hence, we expect compressive strain to strengthen the in-plane interactions and weaken the out-of-plane interactions. The energy scale associated with the magnetic excitations is primarily set by the strength of the in-plane exchange interactions. Thus, we expect compressive strain to increase the value of E$_{(\pi,0)}$. In contrast, the energy scale associated with magnetic ordering is set by the strength of the interactions between neighbouring Ir-O layers, so we expect compressive strain to reduce the size of T$_N$. A similar, but opposite, trend can be expected when tensile strain is applied: stretching the {\it a} and {\it b}-axes will reduce the in-plane interactions and lower E$_{(\pi,0)}$, while shrinking the {\it c}-axis will enhance the out-of-plane interactions and increase $T_N$.

\begin{table}[!ht]
\caption{
Effective singlet-triplet splittings $\Delta E_{\mathrm{ST}}$ in SIO bulk and SIO films for
two adjacent Ir ions (meV).
MRCI+SOC results, see text.
Experimental RIXS values for the zone boundary magnon energy (proportional to $J$)
are also provided.
Strain-induced relative changes in the energy scales are listed in parentheses.
}
%
%\centering % used for centering table
%\setlength{\tabcolsep}{4.5pt}
%
\begin{tabular}{l l l l} % centered columns (5 columns)
\hline %inserts double horizontal lines
Sample     &RIXS: $E_{(\pi,0)}$                     &Model I: $\Delta E_{\mathrm{ST}}$                  &Model II: $\Delta E_{\mathrm{ST}}$ \\[0.5ex]
 %                     &$E_{(\pi,0)}$            &$\Delta E_{\mathrm{ST}}$  &$\Delta E_{\mathrm{ST}}$ \\[0.5ex] 
 % inserts table
%heading
\hline\hline % inserts single horizontal line
SIO-STO    &172\,$\pm$\,4 (--3.4\%)  &49.8 (--12\%)             &56.4 (--0.4\%) \\ % inserting body of the table
SIO-Bulk   &178\,$\pm$\,4            &56.6                      &56.6           \\
SIO-LSAT   &196\,$\pm$\,6 (+10.0\%)  &60.1 (+5.8\%)             &55.0 (--2.8\%) \\ [0.5ex]
%
%$<$ Ir-O-Ir (deg) &156.99 & 156.99 & 156.99 \\
%Ir-Ir (\AA)    &3.9000 & 3.8782 & 3.8700 \\
%Ir-O (\AA) & 1.990, 2.048 & 1.979, 2.057 & 1.975, 2.066  \\ [1ex] % [1ex] adds vertical space
%
\hline %inserts single line
\end{tabular}
\label{table1}
 % is used to refer this table in the text
\end{table}

This argument is supported by {\it ab initio} multireference configuration-interaction (MRCI)
calculations \cite{book_QC_00} on embedded clusters of two nearest-neighbour IrO$_6$ octahedra.
The {\it ab initio} wave function approach has been shown to yield results in good agreement with
experiments measuring the magnetic interactions in $3d$ \cite{CuO_J_Illas_00,Chemrev_Malrieu2014} and $5d$
\cite{Katukuri2012} oxides, as well as determining the dependence of the effective coupling constants on strain
in cuprates \cite{CuO_J_minola_13} and on additional distortions in a few other $d$-metal
compounds \cite{V2O5_J_hozoi_03,CuO_J_bordas_05,Katukuri2012}.
Singlet-triplet splittings for cluster models of bulk SIO and strained films of SIO are listed
in Table~\ref{table1}.
The results were obtained by MRCI calculations including SOC's (MRCI+SOC) \cite{SOC_molpro}.
Two different structural models were employed for the SIO films.
Since the precise structural details are experimentally difficult to access in the films, we assumed
in a first set of calculations that the Ir-O-Ir bond angles are the same as in bulk \cite{Crawford1994}
and only the interatomic distances change with strain (Model I).
At the other extreme, we considered a structural model for which the in-plane Ir-O bond lengths are
fixed to the values measured in bulk \cite{Crawford1994} and for reproducing the strain induced
variation of the lattice parameters \cite{Nichols2013} we modified the Ir-O-Ir angles (Model II).
For Model I, the variations of the average energy of the triplet terms with respect to the singlet
state, denoted in Table I as $\Delta E_{\mathrm{ST}}$, are large, 3.5 to 7 meV.
Since the structure of the triplet components is always the same, with two of them nearly degenerate
and the splitting between the lowest and highest triplet terms taking values in a narrow interval
between 0.9 to 1.2 meV (see the Supplemental Material~\cite{supplmat}), we can safely conclude that for Model I
the most important changes with strain concern the variation of the isotropic Heisenberg exchange $J$.
The overall trend observed in the RIXS spectra for $J$ is in this case nicely reproduced.
In contrast, for Model II, the variations of $\Delta E_{\mathrm{ST}}$ are much smaller and do
not follow the trend observed for $J$ by RIXS.
This suggests that the most significant structural change that occurs in the epitaxial thin
films is the tuning of the Ir-Ir and Ir-O bond lengths. Further detailed investigations of the local structure by local probes such as EXAFS would be extremely useful.

%%%%%%%%%%%%%%%%%%%%%%%%%%
%% Conclusion           %%
%%%%%%%%%%%%%%%%%%%%%%%%%%

In conclusion, we have explored the magnetic properties of $\mathrm{Sr_2IrO_4}$ thin films, with tensile (STO) and compressive (LSAT) epitaxial strain, using x-ray resonant magnetic  scattering and resonant inelastic x-ray scattering. The films show a quasi-two-dimensional magnetic order for both substrates, with magnetic ordering vectors reminiscent of the bulk magnetic structure. Compared to bulk $\mathrm{Sr_2IrO_4}$, the film magnetic correlation lengths show a large anisotropy, with in-plane correlation lengths of $\mathrm{300-400\:\AA}$
%% , compared to very
and very
small $\mathrm{(10-20\:\AA)}$ correlation lengths along the c-axis. We have observed that the magnetic ordering temperature $\mathrm{T_N}$ is suppressed for the compressive strain (LSAT) and enhanced for the tensile strain (STO). In contrast, the RIXS experiments show that the magnetic exchange interactions, determined from the zone-boundary magnon energy, increases (decreases) under compressive (tensile) strain. These results are supported by quantum chemistry calculations, which suggest that the most significant structural change taking place in the films is a tuning of the Ir-O bond length. Other applied perturbations in $\mathrm{Sr_2IrO_4}$ seem to change the magnetic structure of $\mathrm{Sr_2IrO_4}$ at fairly moderate levels (i.e., relatively low applied fields and dopant concentrations). In comparison, a significant epitaxial-strain has no effect on the magnetic ordering wave-vector, while it is altering the energy scales associated with $\mathrm{T_N}$ and J. This illustrates that epitaxial strain is an excellent knob for studying the magnetic properties of iridates.

%

%%%%%%%%%%%%%%%%%%%%%%%%%%
%% Acknowledgement      %%
%%%%%%%%%%%%%%%%%%%%%%%%%%
Research at the U. of Toronto was supported by the NSERC, CFI, and OMRI. Use of the APS was supported by the U. S. DOE, Office of Science, Office of BES, under Contract No. DE-AC02-06CH11357.

\bibliography{iridates_arxiv.bib}

\end{document}